\documentstyle[epsfig,aps,prl,multicol]{revtex}

\begin{document}
\draft
\title{Two-dimensional algorithm of the density matrix
renormalization group}

\author{T. XIANG, J. LOU, and Z. B. SU}

\address{Institute of Theoretical Physics,
Academia Sinica, P. O. Box 2735,
Beijing 100080, The People's Republic of China}

\date{\today}

\maketitle

\begin{abstract}
We propose a new approach to implement the density matrix
renormalization group (DMRG) in two dimensions. With this approach
the initial blocks of a $L\times L$ lattice are built up directly
from the matrix elements of a $(L-1)\times (L-1)$ lattice and the
topological characteristics of two dimensional lattices is preserved
in the iteration of DMRG. By applying it to the spin-1/2
Heisenberg model on both square and triangle lattices, we find that
this approach is significantly more efficient and accurate than other
two-dimensional DMRG methods currently in use.
\end{abstract}

\begin{multicols}{2}

\section{Introduction}

The density matrix renormalization group (DMRG) is an optimized iterative
numerical method. Since its development by White in 1992\cite{White92}, this
method has achieved tremendous success in studying ground state properties
of one-dimensional (1D) interacting electrons. It has also been successfully
extended to finite temperatures\cite{Nishino95,Bursill96}, to momentum space
\cite{Xiang96}, and to the calculation of dynamic correlation functions\cite
{Hallberg95,Anusooya97,Kuhner99}.

The DMRG starts from a small system which can be handled rigourously. A
large chain, called superblock, is then built up from this small system by
adding a number of sites at a time. At each stage, the superblock consists
of system and environment blocks in addition to a number of extra sites.
Graphically, a superblock can be represented as $\left( S\bullet _s\bullet
_eE\right) $, where $S$ and $E$ represent the system and environment blocks
and $\bullet _s$ and $\bullet _e$ the extra sites added to $S$ and $E$,
respectively. $S$ and $\bullet _s$ (similarly $E$ and $\bullet _e$) form an
augmented block, which becomes the system (environment) block in the next
iteration. However, in order to keep the size of the superblock basis from
growing, the basis for the augmented blocks is truncated. Hence the DMRG is a
basis truncation method. However, unlike the conventional renormalization
group method, the truncation is done for each augmented subblock and the
basis states retained are determined not by their energies but by their
probabilities projecting onto the ground state (or other targeted states) of
the superblock. These probabilities are determined by the reduced density
matrix of the augmented system (or environment) block.

To construct the density matrix, the ground state $\left| \psi
\right\rangle $ of the superblock is first diagonalized with the
Lanczos or other sparse matrix diagonalization algorithm. The
reduced density matrix of the augmented system (or environment) is
defined by tracing out from $\left| \psi \right\rangle
\left\langle \psi \right| $ all the degrees of freedom that do not
belong to this block:
\begin{equation}
\rho =Tr_{(E\oplus \bullet _e)}\left| \psi \right\rangle \left\langle \psi
\right| .
\end{equation}
Thus $(E\oplus \bullet _e)$ is considered as a statistical bath to
the augmented system. The density matrix is semi-positive
definite. Its eigenvalue is equal to the projection probability of
the corresponding eigenvector in $\left| \psi \right\rangle $,
i.e.

\begin{equation}
\lambda_l=\sum_j\left| \left\langle \lambda _l,e_j|\psi \right\rangle
\right| ^2,
\end{equation}
where $\left( \lambda _l,\left| \lambda _l\right\rangle \right) $ is an
eigenpair of $\rho $ and $\left\{ \left| e_j\right\rangle \right\} $ is a
basis set of $(E\oplus \bullet _e)$.

Given the density matrix, an entropy can be defined for the augmented system
according to the standard thermodynamic relation
\begin{equation}
S=-Tr\rho \ln \rho=-\sum_l\lambda _l\ln \lambda _l .
\end{equation}
The maximum of the function $f(\lambda )\equiv -\lambda \ln
\lambda $ is located at $\lambda =e^{-1}$. When $0\leq \lambda
<e^{-1}$, $f(\lambda )$ increases monotonically with $\lambda $.
When $\lambda >e^{-1}$, $f(\lambda )$ decreases with $\lambda $.
No more than two $\lambda _l$ can be larger than $e^{-1}$ since
$\sum_l\lambda _l=1$. Thus if the contribution to the entropy 
from the largest $\lambda_l$ is larger than that from the largest 
discarded eigenvalue of $\rho$, the DMRG is also a maximum 
entropy method.

There are two approaches in forming a superblock. In literature
they are often referred as the finite and infinite lattice
approaches. In the infinite lattice approach in one dimension, the
environment block is generally chosen as the space reflection of
the system. In the finite lattice approach, the size of the
superblock is fixed and the environment block is chosen as the
remaining part of the lattice for a given system block. The
infinite lattice approach allows the size of the superblock to be
flexible and can be used to study the thermodynamic limit
directly. However, the finite size approach is more accurate in
calculating quantities for a system with fixed lattice size.

The DMRG can also be used to study thermodynamic properties of a 1D quantum
\cite{Bursill96} or 2D classical system\cite{Nishino95}. In this case, the
transfer matrix of a Hamiltonian system, instead of the Hamiltonian itself,
is diagonalized. The free energy and other thermodynamic quantities are
determined by the maximum eigenvalue of the transfer matrix. The
transfer-matrix DMRG method treats directly an infinite lattice system and
has therefore no finite lattice size effect.

A simple extension of the DMRG to more than 1D would be to replace the
single sites added between the blocks with a row of sites, either along a
principal axis\cite{Jongh98} or along a diagonal\cite{Henelius99}. However,
the extra degrees of freedom added to the system would make the size of the
Hilbert space prohibitively large. Therefore, the two-dimensional algorithm
should be developed so that only a single site is added to each subblock at
a time.

In practice the extension of the DMRG to more than 1D is to map a
higher dimensional lattice onto a 1D one, namely to choose a path
to order all lattice sites\cite{Liang94}. The mapping breaks the
lattice symmetry and introduces long range interactions among
lattice sites. Therefore, the 2D procedure differs from the 1D one
in that there are additional connections between the system and
environment blocks.

A typical mapping, as illustrated in Fig. 1, is to fold a 1D zipper into
2D. This is basically a multi-chain approach since the length of the
folded zipper is unlimited but the width is fixed. For a 2D gas of
non-interacting electrons, Liang and Pang found that the number of states
needed to maintain a certain accuracy grows exponentially with the width of
the lattice \cite{Liang94}. This convergence was also confirmed for an
algorithm where a row of sites was added at each step\cite{Jongh98}.
Although no proof has been given, this statement is often referred to as
most probably valid for any 2D DMRG calculation.

\begin{figure}[htb]
\begin{center}
\epsfig{file=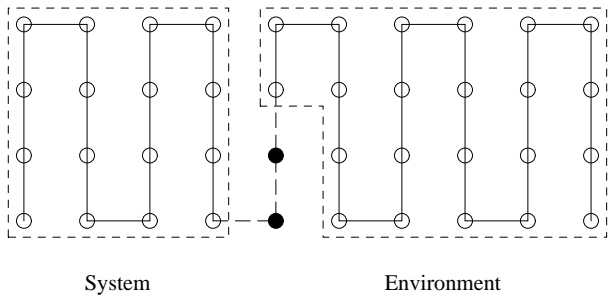,clip=,width=85mm,angle=0}
\end{center}
\caption{A superblock in a ``multichain" algorithm.
The system and environment blocks are enclosed by dashed lines.
Black spots are the sites added to the system and environment. }
\label{manychain}
\end{figure}

This multi-chain approach is simple to implement in the DMRG
iteration. However, with this approach, the calculations on
$(L-1)\times (L-1)$ and $ L\times L$ are performed independently.
The information obtained from the iterations on a $(L-1)\times
(L-1)$ lattice is not used in the preparation of the initial
sub-block matrices in the calculation for a $L\times L$ lattice.
This is undoubtedly a loss of the efficiency. It may result in the
loss of the accuracy as well, since the topological
characteristics of square lattices is not well manifested in the
preparation of the initial block states and the sweeping procedure
of DMRG iterations.

The momentum space DMRG provides an alternative way to implement
the DMRG in two or higher dimensions\cite{Xiang96}. In this
representation the momentum is conserved. This leads to a strong
restriction on the basis states and allows the number of states
kept to increase substantially. Unlike its real space counterpart,
the momentum space DMRG treats the kinetic energy rigorously.
Hence this method works better in the weak coupling limit.
However, the application of the momentum space DMRG has its own
limitations. For example, it is very difficult, if not completely
impossible, to apply this method to a pure spin system like the
Heisenberg model.

In this paper, we introduce a new approach to implement the DMRG in real
space in 2D. Instead of ordering the lattice sites row by row as in the
multi-chain approach, we order the lattice sites by the order along the
diagonal direction. As will be shown later, this is a truly two-dimensional
method which allows us to build up the initial system and environment of a $
L\times L$ lattice system based on the results on a $(L-1)\times (L-1)$
lattice and is particularly suitable for handling 2D lattice models.

The rest of the paper is arranged as the following. In Sec. II a truly 2D
algorithm of the DMRG is introduced. In Sec. III, as an example of the
application of the algorithm, the ground state energy of the spin-1/2
Heisenberg model is evaluated on both square and triangle lattices. The
study is summarized in Sec. IV.

\section{A 2d algorithm of the DMRG}

\begin{figure}[htb]
\begin{center}
\epsfig{file=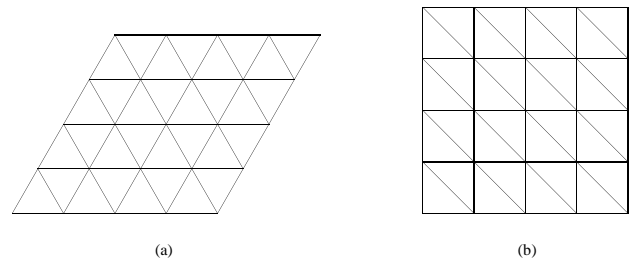,clip=,width=85mm,angle=0}
\end{center}
\caption{A $L\times L$ ($L=5$ here) triangle lattice (a) can be taken as a $L\times L$
square lattice with extra next-nearest neighbor coupling (b). }
\label{triangle}
\end{figure}

In this section we will take the square lattice as an example to
show how to build up initial blocks of a $L\times L$ lattice from a $
(L-1)\times (L-1)$ lattice. The extension to any 2D lattice which can be
topologically transformed to a square lattice by adding or removing some of
the nearest or next nearest neighbor interactions from the square lattice,
such as triangle, hexagonal and Kagomi lattices (an example for such a
transformation is given in Fig. \ref{triangle}), is straightforward.

\begin{figure}[htb]
\begin{center}
\epsfig{file=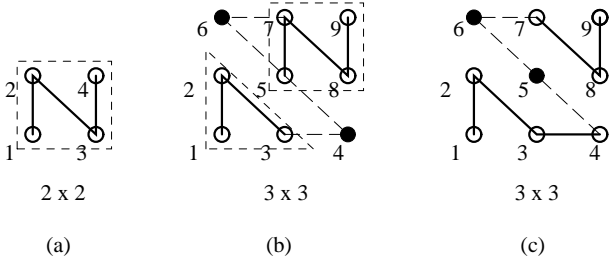,clip=,width=85mm,angle=0}
\end{center}
\caption{(a) is a $2\times 2$ lattice. (b) is the initial configuration 
of superblock for a $3\times 3$ lattice system. 
As indicated by the numbers, the lattice sites are ordered
along the diagonal direction. The initial system contains three sites 
linked by the solid line in the lower left corner. The initial 
environment contains four sites, also linked by the solid line, at 
the upper right corner. (c) same as for (b) but for the next 
iteration. Black spots are the extra sites   
added into the superblock. } \label{fig3by3}
\end{figure}

Let us start from a $2\times 2$ lattice. Fig. \ref{fig3by3}a shows the order
of the sites after the $2D\rightarrow 1D$ mapping. As the system is small,
the Hamiltonian can be fully diagonalized.

Fig. \ref{fig3by3}(b) shows the configuration of the initial superblock for a
$3\times 3$ lattice system. As indicated by the number shown in the figure,
the lattice sites are ordered from the 
lower left corner to the uppper right corner along the diagonal.  
The initial system contains three sites linked by the solid line
in the lower left corner. The initial environment
contains all the four sites in the upper right $2\times 2$
lattice. All the matrix elements for these initial subblocks can
be obtained from the results previously obtained on the $2\times
2$ lattice. We add site $4$ to the system and site $6$ to the
environment to form the augmented system and environment blocks.
Unlike in a real 1D system, these two added sites are not nearest
neighbors in the mapped 1D system. After a standard DMRG
calculation for this superblock, the augmented system block can be
updated and taken as the new system in the next iteration.

In the next iteration (Fig. \ref{fig3by3}c) the system contains four sites 
(i.e. sites 1-4) and the environment contains only three sites at the upper 
right corner (i.e. sites 7-9). Since the two sites (i.e. sites 5 and 6) 
to be added to the system and environment
are nearest neighbors in the mapped 1D lattice, from
now on the DMRG finite system sweeping can be done exactly as in a true 1D system.

\begin{figure}[htb]
\begin{center}
\epsfig{file=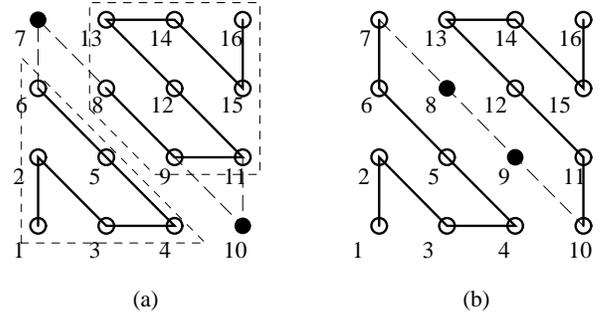,clip=,width=85mm,angle=0}
\end{center}
\caption{(a) a $4\times 4$ lattice decomposed as two partially overlapped
$3\times 3$ lattices (enclosed by the short dashed squares) and two sites
at the two corners outsides these $3\times 3$ lattices
(i.e. sites 6 and 10). The number besides each lattice site gives the order
in the mapped 1D system. The sites in a system or environment block are
linked by solid lines. Black spots are the sites added in.
(b) same as for (a) but for the next iteration.
The environment (sites 10-16) is a space reflection of
the system (sites 1-7) with respect to the center of the $4\times 4$
lattice. }
\label{fig4by4}
\end{figure}

Similarly, the DMRG iterations on a $4\times 4$ lattice can be done based on
the results of the $3\times 3$ lattice. As for a $3\times 3$ lattice, a $
4\times 4$ lattice (Fig. \ref{fig4by4}a) can be formed by two corner cut off 
$3\times 3$ lattices with two isolated sites. The initial system
contains 6 sites linked by a solid line in the lower left corner 
(i.e. sites $1-6$) and the initial environment contains $8$ sites,
also linked by a solid line, in the upper right $3\times 3$ lattice (i.e.
sites $8,9,11-16$). The configurations of these two blocks can be found from
the previously studied $3\times 3$ lattice with or without a space
reflection. We add site $7$ to the system and site $10$ to the environment
to form the augmented system and environment blocks. Again, these two sites
are not nearest neighbors in the mapped 1D system. But the standard DMRG
calculation can be done as usual. The augmented system block is then updated
and taken as the new system in the next iteration.

In the next iteration (Fig. \ref{fig4by4}b), the augmented system in the
last iteration becomes the new system. It contains seven sites (i.e. sites $
1-7$). In this case, since the total number of sites in the environment is
also seven, the environment can therefore be taken as the space reflection
of the system with respect to the center of the $4\times 4$ lattice, i.e.
sites $10-16$. All the matrix elements of this environment can be obtained
from the space reflection of the system. The sites now added into the system
and environment are the two nearest neighboring sites in the mapped 1D
system. Thus starting from this iteration, the standard finite system sweeping can
be done as in a 1D system, without considering how the $4\times 4$ 
lattice is constructed from the $3\times 3$ lattices.

\begin{figure}[htb]
\begin{center}
\epsfig{file=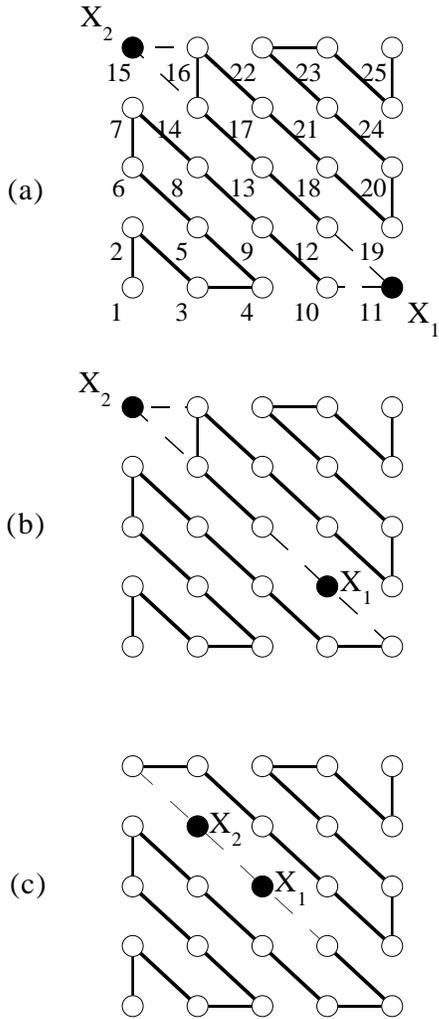,clip=,width=60mm,angle=0}
\end{center}
\caption{The first three initial configurations of superblocks for a 
$5\times 5$ lattice system. At the first two iterations, (a) and (b), 
the two sites which are added to the system and environement (i.e. $X_1$ and $X_2$) 
are not next to each other in the mapped 1D system. At the third iteration, (c),
$X_1$ and $X_2$ become next to each other in the mapped 1D system.}
\label{fig5by5}
\end{figure}

The above procedure can be repeated to larger square lattices. In general, the 
initial superblocks of a $L\times L$ lattice can be formed based on the 
results of the system and environment blocks in a $(L-1)\times (L-1)$ lattice. 
We order all the lattice sites like
a folded zipper with inequal width along the diagonal. If the first site at
the lower left corner of the $L\times L$ lattice is labeled as 1, then the
two sites to be added in will have the coordinates $X_1=(L-1)L/2+1$
and $X_2=L(L+1)/2$ in the mapped 1D system, respectively. 
(An example is given in Fig. \ref{fig5by5} for a $5\times 5$ lattice system.) 
We take the first $ (L-1)L/2$ sites in the lower left corner as the initial
system and all the sites in the upper right $(L-1)\times (L-1)$ square lattice not
used by the system as the initial environment. The DMRG calculation can be
done as before. The system is always augmented and updated. At the first few
iterations, the site which is added to the environment is fixed at $X_2$ and
is not exactly next to $X_1$ in the mapped 1D
lattice. This continues until the environment can be generated by the
center reflection of the system and the two sites added to these two
blocks become nearest neighbors in the mapped 1D system. After that the
standard finite system sweeping can be done as in an ordinary 1D lattice.

\section{The 2d Heisenberg model}

In this section, we take the spin-1/2 Heisenberg model as an
example to demonstrate how good our approach is compared with the
multi-chain approach. The ground state energies on both square and triangle
lattices are evaluated. For these 2D systems, there are currently rather
precise results available, mainly from large-scale Monte Carlo calculations
and series expansions. Therefore the accuracy of our results can be
assessed by comparing with these results.

The Heisenberg model is defined by the Hamiltonian
\begin{equation}
H=\sum_{\left\langle ij\right\rangle }{\bf S}_i\cdot {\bf S}_j
\end{equation}
where ${\bf S}_i$ is the spin operator and the summation runs over all
nearest neighbors. In real space at the same parameters and number of
states, the truncation error in a system with periodic boundary conditions
is usually much higher than with free boundary conditions, therefore we use
free boundary conditions. 

The total spin ${\bf S}^2$ is a good quantum number for the isotropic 
Heisenberg model. This symmetry has been used in obtaining all the results 
presented below. We have also performed finite system iterations using 
both our algorithm and the multichain one. In the multichain 
calculations, we have used an algorithm introduced in Ref \onlinecite{Noack} 
to build up the initial system or environment blocks. 

Table I compares the ground state energy per bond obtained by the true 2D
approach, $E_{2d}$, with that obtained by the multichain approach, $E_{mc}$,
on both square and triangle lattices. For square lattices, $E_{2d}$ is
always lower than $E_{mc}$. Since the DMRG satisfies the variational
principle, this means that the true 2D results are more accurate than the
multichain ones. Moreover, the difference $\left( E_{mc}-E_{2d}\right)
/\left| E_{mc}\right| $ increases with increasing lattice. Thus the
improvement of the true 2D approach over the multichain approach becomes
more and more significant as the lattice size is increased. For triangle
lattices, $E_{2d}$ is slightly higher than $E_{mc}$ when $L$ is small.
However, for large lattices $E_{2d}$ is much more accurate than $E_{mc}$.
The increase of $\left( E_{mc}-E_{2d}\right) /\left| E_{mc}\right| $ with
increasing size in the triangle lattice is even larger than in the square
one.

\begin{figure}[htb]
\begin{center}
\epsfig{file=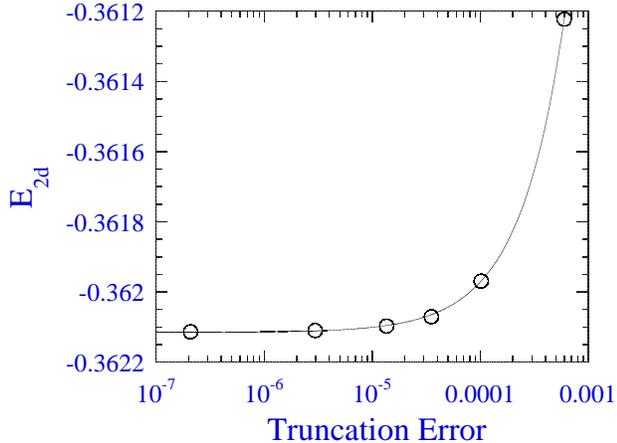,clip=,width=85mm,angle=0}
\end{center}
\caption{The ground state energy per bond
$E_{2d}$ as a function of the truncation error for
the spin-1/2 Heisenberg model on a $6\times 6$
square lattice with free boundary conditions.
The solid line is a polynomial fit to the data.  }
\label{sqtrunerr}
\end{figure}

For a given $L$, an accurate estimate of the ground state energy (similarly
other physical quantities) can be obtained by extrapolating $E_{2d}$ to the
limit $m\rightarrow \infty $. This can also be done\cite{Jeckelmann} 
by extrapolating $E_{2d}$
with respect to the truncation error $\Delta \varepsilon $, since the limit $
m\rightarrow \infty $ is equivalent to the limit $\Delta \varepsilon
\rightarrow 0$. The extrapolation with respect to the number of retained
states is difficult to implement since the asymptotic behavior of $E_{2d}$
in the limit $m\rightarrow \infty $ is unknown and there is some uncertainty
in determining the function used in the extrapolation. However, we find that
the $\Delta \varepsilon $ dependence of $E_{2d}$ is generally very simple
and can be well described by a power law in the limit $\Delta \varepsilon
\rightarrow 0$. An example is given in Fig. \ref{sqtrunerr} where the $
\Delta \varepsilon $ dependence of $E_{2d}$ on a $6\times 6$ square lattice
is shown. In the figure, the solid line is a polynomial fit (up to the
quadratic term in $\Delta \varepsilon $) to the data. From the fit the
ground state energy per bond for this $6\times 6$ system is estimated to be $
-0.36212$. For other cases, this fitting procedure can be similarly done.

To obtain the ground state energy in the thermodynamic limit, we need to do
a finite size scaling for the results obtained from the above extrapolation.
In a periodic system, the leading size correction to the ground state energy
per bond is of order $1/L^3$.\cite{Einarsson95} However, in an open system
as considered here, the finite size effect is stronger and the leading size
correction is of order $1/L$.

\begin{figure}[htb]
\begin{center}
\epsfig{file=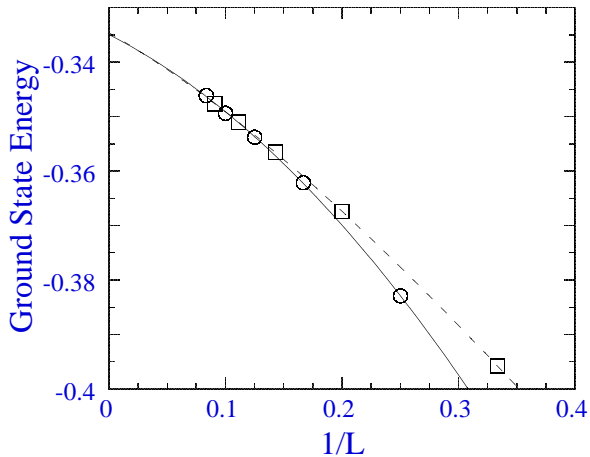,clip=,width=85mm,angle=0}
\end{center}
\caption{Ground state energy $E_{2d}$ versus $1/L$
of the Heisenberg model with free boundary conditions on
square lattices. The behavior of $E_{2d}$ on even lattices is different to
that on odd lattices. But the extrapolated value in the limit $1/L\rightarrow 0$
is the same within numerical errors. }
\label{sqeng}
\end{figure}

\begin{figure}[htb]
\begin{center}
\epsfig{file=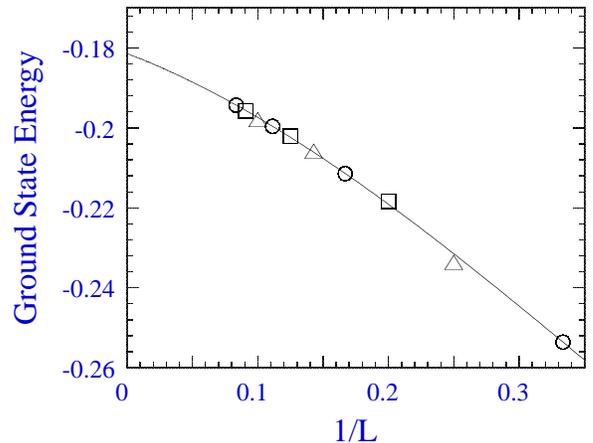,clip=,width=85mm,angle=0}
\end{center}
\caption{$E_{2d}$ versus $1/L$ for
the Heisenberg model with free boundary conditions on
triangle lattices. The solid line is a polynomial fit to
the data with $mod(L,3)=0$. }
\label{trieng}
\end{figure}

Figs. (\ref{sqeng}) and (\ref{trieng}) show the scaling behavior of the
ground state energy on square and triangle lattices, respectively. For the
square lattice, the extrapolated ground state energy in the limit $
1/L\rightarrow 0$ is $E_\infty \approx -0.3346$. This agrees very well with
the probably best currently available estimate, obtained from large-scale
quantum Monte Carlo calculations, of $E_\infty \approx -0.334719(3)$.\cite
{Sandvik97} The result of spin wave theory is $E_\infty =-0.33475$ up to
the fourth order correction\cite{Zheng93}. For the triangle lattice,
the extrapolated result is $E_\infty \approx -0.1814$. It
is also consistent with the quantum Monte Carlo results obtained by
Capriotti et al $\cite{Capriotti99}$, $E_\infty \approx -0.1819$, and by
Bernu et al$\cite{Bernu92}$, $E_\infty \approx -0.1825$. The second order
spin wave result is $E_\infty =-0.1822$.\cite{Miyake92}

The above comparison indicates that accurate results for the ground state
energy can be obtained using the algorithm introduced above. In obtaining
these results, the symmetry of the total spin ${\bf S}^2$ is considered and
up to $300$ states are retained. This calculation can be readily done on a
moderate workstation. With the aid of modern parallel computers, we should
be able to keep more number of states (e.g. $3000$ states) and to further
increase the accuracy.

\section{Conclusion}

We have developed a new approach to implement the real space DMRG in
2D. We point out that a $L\times L$ lattice can be taken as an assembly
of two partially overlapped $(L-1)\times (L-1)$ lattices plus two extra sites
and therefore  the initial blocks of a $L\times L$ system can be built up directly
from the blocks of a $\left( L-1\right) \times \left( L-1\right)$ system.
This is a truly 2D algorithm of the DMRG. It preserves a higher degree of
the symmetry of 2D lattice than the multichain approach and can be readily
used in the DMRG calculation. For
the spin-1/2 Heisenberg model on both square and triangle
lattices, the ground state energies obtained with this approach are consistent with
the quantum Monte Carlo results and better
than those obtained with the multichain approach for large lattice systems.

TX acknowledges the hospitality of the Interdisciplinary Research Center in
Superconductivity of the University of Cambridge, where part of the work was done,
and the financial support from the National Natural Science
Foundation of China and from the special funds for Major State Basic
Research Projects of China.

\end{multicols}

\begin{table}
\caption{Comparison of the ground state energy per bond of the
Heisenberg model on square and triangle lattices
with free boundary conditions obtained by the true
2D approach, $E_{2d}$, with that obtained by the multichain approach,
$E_{mc}$. $m$ is the number of states retained. The lattice size is
$L^2$.}
\begin{tabular}{rccc|ccc}
& \multicolumn{3}{c|}{Square Lattice} & \multicolumn{3}{c}{Triangle Lattice}
\\ \hline
L & $E_{2d}$ & $E_{mc}$ & $(E_{mc}-E_{2d})/\left| E_{mc}\right| $ & E$
_{2d}$ & E$_{mc}$ & $(E_{mc}-E_{2d})/\left| E_{mc}\right|$ \\ \hline
& \multicolumn{6}{c}{$m=50$} \\ \hline
6 & -0.361972 & -0.361919 & 1.5$\times 10^{-4}$ & -0.210692 & -0.210732 &
-1.9$\times 10^{-4}$ \\
8 & -0.35204 & -0.351149 & 2.6$\times 10^{-3}$ & -0.199179 & -0.198752 & 2.1$
\times 10^{-3}$ \\
10 & -0.344292 & -0.341389 & 8.4$\times 10^{-3}$ & -0.192918 & -0.189763 &
1.6$\times 10^{-2}$ \\
12 & -0.337374 & -0.332574 & 1.4$\times 10^{-2}$ & -0.187242 &  -0.182806 &
2.4$\times 10^{-2}$  \\ \hline
& \multicolumn{6}{c}{$m=100$} \\ \hline
6 & -0.362096 & -0.362089 & 1.9$\times 10^{-5}$ & -0.211171 & -0.211196 &
-1.2$\times 10^{-4}$ \\
8 & -0.353213 & -0.353057 & 4.3$\times 10^{-4}$ & -0.200426 & -0.200494 &
-3.3$\times 10^{-4}$ \\
10 & -0.347043 & -0.345771 & 1.3$\times 10^{-3}$ & -0.195015 & -0.192714 &
1.2$\times 10^{-2}$ \\
12 & -0.341588 & -0.338833 & 8$\times 10^{-3}$ & -0.189992 & -0.186441 & 1.9$
\times 10^{-2}$
\end{tabular}
\end{table}

\end{document}